\documentclass[a4paper,english,manuscript]{revtex4}
\usepackage[T1]{fontenc}
\usepackage[latin1]{inputenc}
\usepackage{amsmath}
\usepackage{amssymb}
\usepackage{esint}

\makeatletter

\@ifundefined{textcolor}{}
{%
 \definecolor{BLACK}{gray}{0}
 \definecolor{WHITE}{gray}{1}
 \definecolor{RED}{rgb}{1,0,0}
 \definecolor{GREEN}{rgb}{0,1,0}
 \definecolor{BLUE}{rgb}{0,0,1}
 \definecolor{CYAN}{cmyk}{1,0,0,0}
 \definecolor{MAGENTA}{cmyk}{0,1,0,0}
 \definecolor{YELLOW}{cmyk}{0,0,1,0}
}

\usepackage{babel}
\addto\shorthandsspanish{\spanishdeactivate{~<>}}

\usepackage{babel}

\makeatother

\usepackage{babel}
\begin{document}

\title{Benedicks effect in a relativistic simple fluid }

\author{A. L. Garcia-Perciante$^{1}$, A. Sandoval-Villalbazo$^{2}$, L.
S. Garcia-Colin$^{3}$}

\address{$^{1}$Depto. de Matematicas Aplicadas y Sistemas, Universidad Autonoma
Metropolitana-Cuajimalpa, Artificios 40 Mexico D.F 01120, Mexico.}

\address{$^{2}$Depto. de Fisica y Matematicas, Universidad Iberoamericana,
Prolongacion Paseo de la Reforma 880, Mexico D. F. 01219, Mexico.}

\address{$^{3}$Depto. de Fisica, Universidad Autonoma Metropolitana-Iztapalapa,
Av. Purisima y Michoacan S/N, Mexico D. F. 09340, Mexico. Also at
El Colegio Nacional, Luis Gonzalez Obregon 23, Centro Historico, Mexico
D. F. 06020, Mexico.\\
 }
\begin{abstract}
According to standard thermophysical theories, cross effects are mostly
present in multicomponent systems. In this paper we show that for
relativistic fluids an electric field generates a heat flux even in
the single component case. In the non-relativistic limit the effect
vanishes and Fourier's law is recovered. This result is novel and
may have applications in the transport properties of very hot plasmas. 
\end{abstract}
\maketitle

\section{Introduction}

It is well known that when one considers an ionized plasma as a binary
mixture in the presence of an electromagnetic field, direct and cross
effects arise. Of special interest, are the thermoelectric effects
which have been discussed in Ref. \cite{JNET}. The Benedicks effect,
also called electrothermic effect, corresponds to a heat flux caused
by an electric field. It\textquoteright{}s history can be traced back
to the experimental works developed by Carl Axel Fredrik Benedicks
in the second decade of the twentieth Century in homogeneous systems
\cite{benedicks}. These experiments showed the existence of the reciprocal
cross effect of the Thomson (thermoelectric) effect (1856) which consists
of the generation of a flow of charge due to a temperature gradient.
The Benedicks effect should not be confused with the Peltier effect,
which is strictly related to inhomogeneous systems.

Here, we present what we consider an interesting result which may
lead to applications. If one takes a single component charged relativistic
gas subject to the presence of an electric field, in addition to the
electric current a heat flux appears driven by such field. This effect
is absent in the non-relativistic domain and suggests the possibility
of having additional cross-like effects in relativistic fluids. This
effect may also impact the dynamics of hot plasmas possibly affecting
stability and other properties of such systems.

Several authors have addressed the problem of heat conduction in charged
fluids. In Ref. \cite{kremer ohm} a calculation of the thermoelectric
effect, together with the electric current associated with a temperature
gradient is performed using Anderson and Witting's collisional model
for a binary mixture. Also, in the non-relativistic scenario a reciprocal
effect giving the electrical current due to a temperature gradient
is usually obtained \emph{assuming stationary state} in Boltzmann's
equation \cite{mckelvey}, a procedure that will be carefully discussed
in the last section of the present work.

In this work we establish the relativistic heat flux for a simple
charged fluid in the presence of an electric field starting from a
relativistic Boltzmann equation using Marle's relaxation time approach
to describe the collisional effects. To accomplish this task, a correction
to the relativistic local equilibrium distribution function is obtained
in Sect. II in terms of the gradients in the system to first order
in the Knudsen parameter. Section III is devoted to the explicit calculation
of the heat flux and the corresponding transport coefficients. The
interpretation and discussion of these results are given in Section
IV together with final remarks.

\section{The relativistic boltzmann equation}

We start the calculation by setting up the kinetic equation for the
system in consideration which is a dilute gas of non-degenerate charged
particles at a temperature high enough so that $z=\frac{kT}{mc^{2}}$
is at least close to unity and relativistic effects in the individual
dynamics are relevant. The system as a whole has a hydrodynamic four
velocity which we denote by $\mathcal{U}^{\mu}$ and its constituents
have charge $q$ and mass $m$. The relativistic Boltzmann equation
in the relaxation time approximation reads \cite{ck} 
\begin{equation}
v^{\alpha}\frac{\partial f}{\partial x^{\alpha}}+\dot{v}^{\alpha}\frac{\partial f}{\partial v^{\alpha}}=-\frac{f-f^{\left(0\right)}}{\tau}\label{eq:1}
\end{equation}
The structure of the BGK-like collision kernel was first suggested
by Marle and is considered here as an adequate approximation since
it will lead to the general structure of the distribution function
as well as the dissipative fluxes without introducing all the complicated
expressions and calculations that are involved when considering the
complete collision kernel. The complete calculation would perhaps
yield relevant corrections to the particular values of the transport
coefficients but both the structure of the fluxes as well as the general
behavior of the coefficients will remain unaltered. Here $f$ is the
distribution function, $f^{\left(0\right)}$ the local equilibrium
solution, and $\tau$ a characteristic relaxation time. The molecular
dynamic variables are the individual particle four-velocity $v^{\mu}$
and the corresponding acceleration $\dot{v}^{\mu}$, both referred
to the laboratory frame.

Since the fluid is charged, in presence of an electrostatic field
the acceleration in the second term on the left side of Eq. (\ref{eq:1})
may be written as 
\begin{equation}
\dot{v}^{\mu}=\frac{q}{m}v_{\alpha}F^{\alpha\mu}\label{eq:2}
\end{equation}
 where 
\begin{equation}
F^{\alpha\mu}=\left(\begin{array}{cccc}
0 & 0 & 0 & -\frac{\phi^{,1}}{c}\\
0 & 0 & 0 & -\frac{\phi^{,2}}{c}\\
0 & 0 & 0 & -\frac{\phi^{,3}}{c}\\
\frac{\phi^{,1}}{c} & \frac{\phi^{,2}}{c} & \frac{\phi^{,3}}{c} & 0
\end{array}\right)\label{eq:3}
\end{equation}
 is the usual electromagnetic field tensor and thus 
\begin{equation}
m\dot{v}^{\mu}=\begin{cases}
-qv^{4}\left(\frac{\phi^{,\mu}}{c}\right) & \quad\mu=1,2,3\\
m\dot{v}^{4}=-qv_{\ell}\frac{\phi^{,\ell}}{c}\quad & \mu=4
\end{cases}\label{eq:4}
\end{equation}
Here a coma indicates a partial derivative whereas a semicolon will
denote a covariant one. For the special relativistic fluid, we consider
a flat Minkowski spacetime with a $+++-$ signature such that the
covariant derivative will coincide with the ordinary one in cartesian
coordinates. The dot represents a proper time derivative. 

Substitution of Eq. (\ref{eq:2}) in Eq. (\ref{eq:1}) leads to 
\begin{equation}
v^{\alpha}f_{,\alpha}+\frac{q}{m}v_{\alpha}F^{\alpha\mu}\frac{\partial f}{\partial v^{\mu}}=-\frac{f-f^{\left(0\right)}}{\tau}\label{eq:4.5}
\end{equation}
which will be solved, following the standard Chapman-Enskog procedure,
by introducing the assumption that the distribution function can be
written as 
\begin{equation}
f=f^{\left(0\right)}+f^{\left(1\right)}\label{eq:6}
\end{equation}
where $f^{\left(1\right)}$ is of first order in the gradients. By
substituting Eq. (\ref{eq:6}) in Eq. (\ref{eq:4.5}) and keeping
only terms up to first order in the gradients, one can write the deviation
from the equilibrium solution as follows 
\begin{equation}
f^{\left(1\right)}=-\tau\left\{ v^{\alpha}f_{,\alpha}^{\left(0\right)}+\left(\frac{q}{m}v_{\alpha}F^{\alpha\mu}\right)\frac{\partial f^{\left(0\right)}}{\partial v^{\mu}}\right\} \label{eq:7}
\end{equation}
The presence of the local equilibrium distribution function on the
right side of Eq. (\ref{eq:7}) follows from the fact that its derivatives
are already first order in the gradients since the local equilibrium
assumption asserts that its space and time dependence is given only
though the gradients and time derivatives of the state variables $n$,
$T$ and $\mathcal{U}^{\nu}$. It must also be taken into account
that the time derivatives of such variables are in turn written in
terms of gradients by introducing Euler's relativistic equations,
a step necessary in order to assure existence of the first order solution
\cite{courant,ChCow}. 

In order to establish the explicit dependence of $f^{\left(1\right)}$
with the gradients, we start by using the following identity 
\begin{equation}
v^{\alpha}=v^{\beta}h_{\beta}^{\alpha}+\gamma_{\left(k\right)}\mathcal{U}^{\alpha}\label{eq:8}
\end{equation}
where $h_{\beta}^{\alpha}$ projects in the hyperplane orthogonal
to the hydrodynamic velocity, that is $h_{\beta}^{\alpha}\mathcal{U}^{\beta}=0$.
Also, for the last equality use has been made of the identity $v^{\beta}\mathcal{U}_{\beta}=-\gamma_{\left(k\right)}c^{2}$
which is valid since, being the contraction a scalar, it can be calculated
in a comoving frame where $\mathcal{U}^{\nu}=\left[\vec{0},c\right]$
and $v^{\alpha}\equiv K^{\alpha}$, with 
\begin{equation}
K^{\alpha}=\gamma_{\left(k\right)}\left[k^{\ell},c\right]\qquad\ell=1,2,3\label{eq:9}
\end{equation}
denoting the chaotic or peculiar velocity \cite{ck,ChCow} and $\gamma_{\left(k\right)}=\left(1-k^{2}/c^{2}\right)^{-1/2}$
being the usual Lotentz factor with $k=\sqrt{k^{\ell}k_{\ell}}$.
By noticing that $v^{\alpha}f_{,\alpha}^{\left(0\right)}$ is also
an invariant quantity, the calculation is carried out in the comoving
frame where the projector has the simple form 
\begin{equation}
h_{\beta}^{\alpha}=\left(\begin{array}{cccc}
1 & 0 & 0 & 0\\
0 & 1 & 0 & 0\\
0 & 0 & 1 & 0\\
0 & 0 & 0 & 0
\end{array}\right)\label{eq:11}
\end{equation}
so that one can write in general 
\begin{equation}
v^{\alpha}f_{,\alpha}^{\left(0\right)}=\left.v^{\alpha}f_{,\alpha}^{\left(0\right)}\right|_{CF}=\gamma_{\left(k\right)}k^{i}\delta_{i}^{\ell}\left.f_{,\ell}^{\left(0\right)}\right|_{CF}+\gamma_{\left(k\right)}c\left.f_{,4}^{\left(0\right)}\right|_{CF}\label{eq:12}
\end{equation}
where $i,\,\ell=1,2,3$ and the derivatives are calculated in the
comoving frame, denoted by the subscript $CF$, where local equilibrium
is satisfied. Since locally the equilibrium distribution function
is a Juttner function given by \cite{ck,juttner}

\begin{equation}
f^{(0)}=\frac{n}{4\pi c^{3}z\mathcal{K}_{2}\left(\frac{1}{z}\right)}\exp\left(\frac{\mathcal{U}^{\beta}v_{\beta}}{zc^{2}}\right)\label{eq:13}
\end{equation}
one has 
\begin{equation}
\frac{\partial f^{\left(0\right)}}{\partial n}=\frac{f^{\left(0\right)}}{n}\qquad\frac{\partial f^{\left(0\right)}}{\partial\mathcal{U}^{\alpha}}=\frac{v_{\alpha}}{zc^{2}}f^{\left(0\right)}\qquad\frac{\partial f^{\left(0\right)}}{\partial v^{\alpha}}=\frac{\mathcal{U}_{\alpha}}{zc^{2}}f^{\left(0\right)}\label{eq:14}
\end{equation}
and 
\begin{equation}
\frac{\partial f^{\left(0\right)}}{\partial T}=\frac{f^{\left(0\right)}}{T}\left(1-\frac{\gamma_{\left(k\right)}}{z}-\frac{\mathcal{G}\left(\frac{1}{z}\right)}{z}\right)\label{eq:14.5}
\end{equation}
where $\mathcal{G}\left(\frac{1}{z}\right)=\frac{\mathcal{K}_{3}\left(\frac{1}{z}\right)}{\mathcal{K}_{2}\left(\frac{1}{z}\right)}$
and $\mathcal{K}_{n}\left(\frac{1}{z}\right)$ is the modified $n$-th
order Bessel function of the second kind depending on the relativistic
parameter $z=\frac{kT}{mc^{2}}$ . Thus, 
\begin{equation}
\left.f_{,\alpha}^{\left(0\right)}\right|_{CF}=f^{\left(0\right)}\left[\frac{n_{,\alpha}}{n}+\frac{T_{,\alpha}}{T}\left(1-\frac{\gamma_{\left(k\right)}}{z}-\frac{\mathcal{G}\left(\frac{1}{z}\right)}{z}\right)+\frac{v_{\beta}}{zc^{2}}\mathcal{U}_{;\alpha}^{\beta}\right]\label{eq:15}
\end{equation}
and for $f_{,4}^{\left(0\right)}=\frac{1}{c}\frac{\partial f^{\left(0\right)}}{\partial t}$
we have 
\begin{equation}
\left.f_{,4}^{\left(0\right)}\right|_{CF}=f^{\left(0\right)}\left[\frac{1}{n}\left.\frac{\partial n}{\partial t}\right|_{CF}+\frac{1}{T}\left(1-\frac{\gamma_{\left(k\right)}}{z}-\frac{\mathcal{G}\left(\frac{1}{z}\right)}{z}\right)\left.\frac{\partial T}{\partial t}\right|_{CF}+\frac{K_{\beta}}{zc^{2}}\left.\frac{\partial\mathcal{U}^{\beta}}{\partial t}\right|_{CF}\right]\label{eq:16}
\end{equation}
Following Hilbert's procedure \cite{ChCow}, the time derivatives
are to be written in terms of the gradients of the state variables
via Euler's (lower order) equations:

\begin{equation}
\frac{\partial n}{\partial t}+\mathcal{U}^{\ell}n_{,\ell}=-n\mathcal{U}_{;\mu}^{\mu}\label{eq:17}
\end{equation}
 
\begin{equation}
\frac{\partial\mathcal{U}^{\mu}}{\partial t}+\mathcal{U}^{\ell}\mathcal{U}_{;\ell}^{\mu}=-h^{\mu\nu}\frac{nmc^{2}z}{\tilde{\rho}}\left(\frac{T_{,\nu}}{T}+\frac{n_{,\nu}}{n}\right)+n\frac{q}{\tilde{\rho}}\mathcal{U}_{\nu}F^{\nu\mu}\label{eq:18}
\end{equation}
 
\begin{equation}
\frac{\partial T}{\partial t}+\mathcal{U}^{\ell}T_{,\ell}=-\frac{kT}{C_{n}\left(z\right)}\mathcal{U}_{;\alpha}^{\alpha}\label{eq:19}
\end{equation}
where $C_{n}$ is the specific heat for constant $n$ and for and
ideal gas $\tilde{\rho}=\frac{n\varepsilon}{c^{2}}+\frac{p^{2}}{c^{2}}=nm\mathcal{G}\left(\frac{1}{z}\right)$.
The set given by Eqs. (\ref{eq:17})-(\ref{eq:19}) corresponds to
the local equilibrium, zero order in the gradients, relativistic hydrodynamic
equations which can be obtained in the kinetic theory framework by
multiplying Boltzmann's equation by collisional invariants and integrating
in velocity space. Such a procedure can be found for the relativistic
case in Ref. \cite{ck} for example. In order to obtain such relations
in the comoving frame, as required by Eq. (\ref{eq:16}), we consider
$\mathcal{U}^{\nu}=\left[\vec{0},c\right]$ and Eq. (\ref{eq:11})
which leads to 
\begin{equation}
\left.\frac{\partial n}{\partial t}\right|_{CF}=0\label{eq:20}
\end{equation}
 
\begin{equation}
\left.\frac{\partial\mathcal{U}^{\ell}}{\partial t}\right|_{CF}=-\frac{1}{\mathcal{G}\left(\frac{1}{z}\right)}\left[h^{\ell\nu}c^{2}z\left(\frac{T_{,\nu}}{T}+\frac{n_{,\nu}}{n}\right)-\frac{q}{m}cF^{4\ell}\right]\label{eq:21}
\end{equation}
 
\begin{equation}
\left.\frac{\partial\mathcal{U}^{4}}{\partial t}\right|_{CF}=0\label{eq:22}
\end{equation}
 
\begin{equation}
\left.\frac{\partial T}{\partial t}\right|_{CF}=0\label{eq:23}
\end{equation}
Introducing Eqs. (\ref{eq:20}) to (\ref{eq:23}) in Eq. (\ref{eq:16})
one obtains 
\begin{align}
v^{\alpha}f_{,\alpha}^{\left(0\right)} & =f^{\left(0\right)}\gamma_{\left(k\right)}k^{i}h_{i}^{\ell}\left\{ \left(1-\frac{\gamma_{\left(k\right)}}{\mathcal{G}\left(\frac{1}{z}\right)}\right)\frac{n_{,\ell}}{n}+\frac{\gamma_{\left(k\right)}}{\mathcal{G}\left(\frac{1}{z}\right)}\left[\frac{q}{zmc^{2}}cF^{4\ell}\right]\right.\nonumber \\
 & \left.+\frac{T_{,\ell}}{T}\left(1-\frac{\gamma_{\left(k\right)}}{z}-\frac{\gamma_{\left(k\right)}}{\mathcal{G}\left(\frac{1}{z}\right)}-\frac{\mathcal{G}\left(\frac{1}{z}\right)}{z}\right)\right\} \label{eq:24}
\end{align}
where a term proportional to the velocity gradient is ignored from
now on since it will not couple with the heat flux. This fact is sustained
by Curie's principle by means of which only fluxes and forces of the
same tensorial rank are coupled to each other in the constitutive
equations.

On the other hand, the second term in Eq. (\ref{eq:7}) is calculated
using the fact that the field tensor only has electric potential components
($4-\ell$ with $\ell$ running up to 3) such that 
\begin{equation}
\left(\frac{q}{m}v_{\alpha}F^{\alpha\mu}\right)\frac{\mathcal{U}_{\mu}}{zc^{2}}=\frac{q}{m}v_{\alpha}F^{\alpha4}\frac{\mathcal{U}_{4}}{zc^{2}}=\frac{q}{m}v_{\alpha}F^{\alpha4}\frac{\mathcal{U}_{4}}{zc^{2}}\label{eq:25}
\end{equation}
The quantity is scalar and therefore can be calculated in the comoving
frame, so that

\begin{equation}
\left(\frac{q}{m}v_{\alpha}F^{\alpha\mu}\right)\frac{\mathcal{U}_{\mu}}{zc^{2}}=\gamma_{\left(k\right)}\frac{q}{m}k_{\ell}F^{\ell4}\frac{c}{zc^{2}}\label{eq:26}
\end{equation}
and thus, since $F^{4\ell}=\frac{\phi^{,\ell}}{c}=-F^{\ell4}$ 
\begin{align}
f^{\left(1\right)} & =-\tau f^{\left(0\right)}\gamma_{\left(k\right)}h_{i}^{\ell}k^{i}\left\{ \frac{1}{z}\left(\frac{\gamma_{\left(k\right)}}{\mathcal{G}\left(\frac{1}{z}\right)}-1\right)\frac{q}{mc^{2}}\phi_{,\ell}+\left(1-\frac{\gamma_{\left(k\right)}}{\mathcal{G}\left(\frac{1}{z}\right)}\right)\frac{n_{,\ell}}{n}\right.\nonumber \\
 & \left.+\left(1-\frac{\gamma_{\left(k\right)}}{z}-\frac{\gamma_{\left(k\right)}}{\mathcal{G}\left(\frac{1}{z}\right)}-\frac{\mathcal{G}\left(\frac{1}{z}\right)}{z}\right)\frac{T_{,\ell}}{T}\right\} \label{eq:27}
\end{align}
The function in Eq. (\ref{eq:27}) is the correction to the equilibrium
distribution function to first order in the gradients in the presence
of an electric field. This part of the distribution yields the dissipative
terms in the hydrodynamic equations to the Navier-Stokes level. In
the next section, $f^{\left(1\right)}$ will be introduced in the
expression for the heat flux in the relativistic case in order to
establish the general structure of the corresponding constitutive
equation as well as the particular value of the transport coefficients
for the relaxation approximation.

\section{Constitutive equation for the heat flux}

An analysis of both the concept and the formal expression for the
heat flux in relativistic kinetic theory in relation to the flux of
chaotic energy, as initially conceived by R. Clausius \cite{clausius},
can be found in Ref. \cite{JNET}. In this context it can be shown
that the heat flux in the comoving frame is given by 
\begin{equation}
J_{\left[Q\right]}^{\nu}=mc^{2}h_{\mu}^{\nu}\int k^{\mu}f^{\left(1\right)}\gamma_{\left(k\right)}^{2}d^{*}K\label{eq:heatFlux rel-1}
\end{equation}
and can be expressed in an arbitrary frame using a Lorentz transformation.
From (\ref{eq:heatFlux rel-1}) and considering the structure of Eq.
(\ref{eq:27}) it can be seen that three independent driving forces
will be present in the flux namely, the temperature, particle density
and electric potential gradients. Of particular interest in this work
is the term that corresponds to the electric field which in this relaxation
time approximation is given by

\begin{equation}
J_{\left[Q,E\right]}^{\ell}=-\tau\frac{qh_{i}^{\ell}\phi^{,i}}{3z}\int k^{2}f^{\left(0\right)}\left(\frac{\gamma_{\left(k\right)}}{\mathcal{G}\left(\frac{1}{z}\right)}-1\right)\gamma_{\left(k\right)}^{3}d^{*}K,\label{eq:29}
\end{equation}
Notice that this flux vanishes in the non-relativistic limit where
both $\gamma_{k}$ and $\mathcal{G}\left(\frac{1}{z}\right)$ are
unity. Thus, the heat flux in the single component fluid due to the
thermoelectric effect is only present for non-negligible values of
the parameter $z$ and can be written as 
\begin{equation}
J_{\left[Q,E\right]}^{\ell}=L_{TE}h_{i}^{\ell}\phi^{,i}\label{eq:30}
\end{equation}
where the corresponding transport coefficient is given by

\begin{equation}
L_{TE}=n\tau c^{2}q\left\{ 5z+\frac{1}{\mathcal{G}\left(\frac{1}{z}\right)}-\mathcal{G}\left(\frac{1}{z}\right)\right\} \label{eq:31}
\end{equation}
For small $z$, in the non-relativistic limit one has 
\begin{equation}
L_{TE}=n\tau c^{2}q\frac{5}{2}\left(z^{2}-z^{3}+...\right)\label{eq:32}
\end{equation}
or equivalently 
\begin{equation}
L_{TE}\sim\frac{5}{2}n\tau\left(\frac{kT}{m}\right)^{2}\frac{q}{c^{2}}\label{eq:33}
\end{equation}
The general form for the heat flux including all the gradients is
given by 
\begin{equation}
J_{\left[Q\right]}^{\mu}=-h^{\mu\nu}\left(L_{TT}\frac{T_{,\nu}}{T}+L_{Tn}\frac{n_{,\nu}}{n}+L_{TE}\phi_{,\nu}\right)\label{eq:34}
\end{equation}
where the transport coefficients corresponding to the first two terms
were previously obtained \cite{pa2009}. In the non-relativistic limit,
the ordering of the effects with respect to $c$ can be extracted
from the following expression 
\begin{equation}
J_{\left[Q\right]}^{\mu}\sim-\frac{5}{2}n\tau\frac{k^{2}T^{2}}{m}h^{\mu\nu}\left[\frac{T_{,\nu}}{T}+\frac{kT}{mc^{2}}\frac{n_{,\nu}}{n}+\frac{q}{mc^{2}}\phi_{,\nu}\right]\label{eq:35}
\end{equation}
from where it can be clearly seen that the only effect that survives
in the non-relativistic limit is the one corresponding to Fourier's
law and the other two terms are relevant only when the thermal energy
and/or the electric energy are comparable in magnitude to the rest
mass of the molecules. It is worthwhile to remind the reader at this
point that the results obtained in this section are valid for single
component gases, no binary mixture needs to be considered.

\section{Summary and final remarks}

In this paper, we have studied the heat flux in a single component
relativistic charged gas in the presence of an electric field. The
calculation was performed using Marle's kernel for the Boltzmann equation.
The result is a heat flux coupled to three forces, namely the gradients
of the temperature, density and electric potential. Only the temperature
gradient contribution does not vanish in the non-relativistic limit
while the other two effects are strictly relativistic.

Some authors have argued in favor of a cross-like relationship between
temperature gradients and electric field in the non-relativistic single
component fluid \cite{mckelvey}. In order for such an effect to be
established, one would have to set the partial time derivative in
Boltzmann's equation to zero. This is usually justified by assuming
a steady state in the gas, however the Chapman-Enskog procedure arises
from an ordering scheme in Boltzmann's equation in which the time
derivative of the equilibrium distribution function results as a first
order in the gradients term and thus must be considered when calculating
the first order correction. Also, if the Benedicks effect would exist
in the non-relativistic single component fluid in steady state, one
should expect it to be recoverable as a limit of the complete solution.

By inspection of the structure of the heat flux given by Eq. (\ref{eq:34}),
one can draw interesting conclusions regarding the physics of dissipative
simple relativistic charged gases. If for the sake of simplicity one
ignores the particle density for a homogeneous system, the constitutive
equation has a structure similar to the one leading to Tolman's law
in the case of a gravitational field \cite{tolman}. Reasoning in
a similar way as in Ref. \cite{tolman}, if one assumes a vanishing
heat flux, then to lowest order in $z$: 
\begin{equation}
\frac{T_{,\nu}}{T}=-\frac{q}{mc^{2}}\phi_{,\nu}\label{eq:36}
\end{equation}
in such a way that a temperature gradient can be sustained in the
absence of a heat flux by means of an electric field for the simple
charged gas. It is interesting to notice that for the case of negative
and positive charged particles, the required electric potential gradient
reverses direction for the same temperature gradient.

Another interesting question arises here regarding the possibility
of a relativistic version of a Wiedemann-Franz like relation for a
single component fluid. This could lead to experimental tests for
the strictly relativistic effects here presented, as well as for the
pertinent generalizations of this formalism including magnetic fields. 

\textsf{\textbf{\textit{\Large Acknowledgements}}}{\Large \par}

The authors acknowledge support from CONACyT through grant number
167563.

\end{document}